\documentclass[11pt,a4paper,twoside,groupcitations]{article}
\usepackage[T1]{fontenc}
\usepackage[ansinew]{inputenc}
\usepackage[english]{babel}
\usepackage{amsfonts}
\usepackage{amsmath}
\usepackage{bm}
\usepackage{array}
\usepackage{amsthm}
\usepackage{amssymb}
\usepackage{graphicx}
\usepackage{subfigure}
\usepackage{braket}
\usepackage{eucal}
\usepackage{verbatim}
\usepackage[table]{xcolor}
\usepackage{caption}
\usepackage{cite}
\usepackage{textcomp}
\usepackage{multicol}
\usepackage{tikz}
\usetikzlibrary{positioning,arrows}
\usetikzlibrary{decorations.pathmorphing}
\usetikzlibrary{decorations.markings}
\usetikzlibrary{calc,decorations.markings}
\usetikzlibrary{arrows,shapes}
\usetikzlibrary{matrix,arrows}
\usepackage{pgfplots}
\usepackage{xparse}
\definecolor{jade}{HTML}{00A86B}
\newcommand{\be}{\begin{eqnarray}}
\newcommand{\ee}{\end{eqnarray}}

\newcommand{\expec}[1]{\mbox{$\langle\, #1\,\rangle$}}

\renewcommand{\d}{\mbox{${\rm d}$}} 
\newcommand{\lp}{\ell_{\rm p}}
\newcommand{\mpl}{m_{\rm p}}
\newcommand{\gn}{G_{\rm N}}

\newcommand{\Rh}{R_{\rm H}}

\title{\bf A quantum bound on the compactness}
\author{Roberto~Casadio$^{ab}$\thanks{E-mail: casadio@bo.infn.it}
\\
\\
$^a${\em Dipartimento di Fisica e Astronomia, Universit\`a di Bologna}
\\
{\em via Irnerio~46, 40126 Bologna, Italy}
\\
\\
$^b${\em I.N.F.N., Sezione di Bologna, I.S.~FLAG}
\\
{\em viale B.~Pichat~6/2, 40127 Bologna, Italy}
}
\begin{document}
\maketitle
\begin{abstract}
We present a simple quantum description of the gravitational collapse of a ball
of dust which excludes those states whose width is arbitrarily smaller than the gravitational
radius of the matter source and supports the conclusion that black holes are
macroscopic extended objects.
We also comment briefly on the relevance of this result for the ultraviolet
self-completion of gravity and the connection with the corpuscular picture of black holes.
\end{abstract}
\section{Introduction and motivation}
\setcounter{equation}{0}
\label{Sintro}
It has been often argued that quantum gravity should remove the singularity
predicted by General Relativity at the endpoint of the gravitational collapse~\cite{HE}.
We study this issue by considering a simple quantum description of the gravitational
collapse of a ball of dust~\cite{OS}.
Since gravity is the only interaction acting on dust, the areal radius $R$ of the ball
classically follows a radial geodesic in the (outer) Schwarzschild spacetime~\footnote{We
use units with $c=1$, the reduced Planck constant $\hbar=\lp\,\mpl$ and the Newton constant
$\gn=\lp/\mpl$, where $\lp$ is the Planck length and $\mpl$ the Planck mass.}
\be
\d s^2
=
-\left(1-\frac{2\,\gn\,M}{r}\right)
\d t^2
+
\left(1-\frac{2\,\gn\,M}{r}\right)^{-1}
\d r^2
+
r^2\,\d\Omega^2
\ ,
\label{schw}
\ee
where $M$ is the Arnowitt-Deser-Misner (ADM)~\cite{ADM} mass of the dust.
In analogy with the quantum mechanics of the hydrogen atom, in which
one quantises the position of the electron with respect to the
centre-of-mass of the system, only the radius $R$ of the ball will be
quantised here.~\footnote{For more refined approaches in which the singularity
is also removed, see {\em e.g.}~Refs.~\cite{bounce}.}
The radial geodesic equation will then become a time-independent Schr\"odinger
equation for a particle in the Newtonian potential, with the important feature that
quantum states with width significantly smaller than the gravitational radius 
\be
\Rh=2\,\gn\,M
\label{Rh}
\ee
are not physically allowed.
This is similar to the quantum removal of the classical UV catastrophe 
in the hydrogen atom, except that the mass $M$ naturally introduces
a lower bound on the energy spectrum which would be absent in the
Newtonian approximation.
One could then infer that the quantum nature of black holes as
extended objects follows from the non-linearity of the gravitational 
interaction as it is described by General Relativity.
\par
The main results will be obtained in the next Section, with more speculative
considerations summarised in Section~\ref{S:conc}.
\section{Minisuperspace model for the gravitational collapse} 
\label{S:mini}
\setcounter{equation}{0}
We consider a self-gravitating ball of dust of radius $R$ and ADM mass $M$~\cite{OS,stephani}.
In General Relativity, the surface of the ball follows a radial geodesic in the Schwarzschild spacetime~\eqref{schw}
and its areal radius $R=R(\tau)$ must therefore satisfy an equation of the form~\footnote{Numerical coefficients
of order one will often be omitted or approximated for the sake of clarity, given that the simplicity of the model
diminishes their physical relevance.
A more precise analysis is left for future refinements.}
\be
\left(\frac{\d R}{\d \tau}\right)^2
+
1-\frac{2\,\gn\,M}{R}
\simeq
\frac{E^2}{M^2}
\ ,
\ee
where $\tau$ is the proper time and $E<M$ is the conserved energy of a bound trajectory.
The above can be rewritten as
\be
H
\equiv
\frac{P^2}{2\,M}
-
\frac{\gn\,M^2}{R}
=
\frac{M}{2}
\left(
\frac{E^2}{M^2}
-1
\right)
\equiv
\mathcal E
\ ,
\label{radial}
\ee
where we introduced the momentum $P=M\,(\d R/\d \tau)$.
One can immediately notice that Eq.~\eqref{radial} is formally the same as the Newtonian equation
for energy conservation.
In fact, Eq.~\eqref{radial} is the mass-shell condition for the ball and the Hamiltonian constraint
of General Relativity for dust.
\par
The study of spherically symmetric gravitational collapse in the quantum theory has a long history
(for a partial selection of papers, see Refs.~\cite{bounce,qcollapse}).
We here proceed with a simple quantisation prescription for the above equation by introducing
the momentum operator $\hat P=-i\,\hbar\,\partial/\partial R$.
This is physically equivalent to assuming that the radius of the ball $R$ satisfies an uncertainty
relation stemming from the usual canonical commutator, that is
\be
\left[
\hat R,\hat P
\right]
=
i\,\hbar
\quad
\Rightarrow
\quad
\Delta R\,\Delta P
\gtrsim
\hbar
=
\lp\,\mpl
\ ,
\ee
where $\Delta O\equiv\expec{\hat O^2}-\expec{\hat O}^2$ for $\hat O=\hat R$ or $\hat P$.
Moreover, all of the expectation values are taken on wavefunctions $\Psi=\Psi(R)$ satisfying
\be
\hat H\,\Psi
=
\mathcal E\,\Psi
\ ,
\ee
which is just the time-independent Schr\"odinger equation for a gravitational (Newtonian) atom.
In particular, the energy spectrum contains the eigenstates
\be
\Psi_n
\simeq
\sqrt{\frac{M^9}{\pi\,n^5\,\lp^3\,\mpl^9}}\,
e^{-\frac{M^3\,r}{n\,\mpl^3\,\lp}}\,
L_{n-1}^1
\left(\frac{2\,M^3\,r}{n\,\mpl^3\,\lp}
\right)
\ .
\ee
where the integer number $n\ge 1$ and $L_{n-1}^1$ are the generalised Laguerre polynomials (for zero angular 
momentum~\footnote{An obvious generalisation is to consider a spinning ball, but that breaks
spherical symmetry, and a treatment fully consistent with General Relativity would thus become
much more involved.}).
The corresponding eigenvalues are given by
\be
\frac{\mathcal E_n}{M}
\simeq
-
\frac{G_{\rm N}^2\,M^4}{2\,\hbar^2\,n^2}
=
-
\frac{1}{2\,n^2}
\left(\frac{M}{\mpl}\right)^4
=
\frac{1}{2}
\left(
\frac{E_n^2}{M^2}
-1
\right)
\ ,
\label{eqE}
\ee
and one also has
\be
R_n
\equiv
\bra{\Psi_n} R\ket{\Psi_n}
\simeq
\frac{\hbar^2\,n^2}{\gn\,M^3}
=
n^2\,\lp
\left(\frac{\mpl}{M}\right)^3
\ .
\ee
At first sight, it thus appears that the spectrum contains states $\Psi_n$ of infinitesimally small
width, since $R_1\sim \lp\,(\mpl/M)^3\ll \lp$ for any macroscopic objects, like stars, whose mass
$M\gg \mpl$.
This ground state would have an energy density of the order of $M/R_1^3\sim (M^{10}/\mpl^9)\,\lp^{-3}$,
which can hardly be considered a satisfying alternative to the classical singularity of infinte energy density.
\par
However, Eq.~\eqref{eqE} yields
\be
0\le
\frac{E_n^2}{M^2}
\simeq
1
-
\frac{1}{n^2}
\left(\frac{M}{\mpl}\right)^4
\ ,
\ee
tantamount to $\mathcal E\gtrsim -M/2$ in Eq.~\eqref{radial}, and we thus find that
acceptable states $\Psi_n$ must satisfy
\be
n
\ge N_M
\simeq
\left(\frac{M}{\mpl}\right)^2
\ .
\label{eqn0}
\ee
Correspondingly, we obtain 
\be
R_n
\gtrsim
R_{N_M}
\sim
\Rh
\ ,
\label{RnRh}
\ee
which results in the quantum upper bound 
\be
\frac{\gn\,M}{R_n}
\lesssim
1
\ .
\label{eqX}
\ee
Moreover, we observe that the Hamiltonian eigenvalues are also bounded from below as
\be
\mathcal E_n
\ge
\mathcal E_{N_M}
\simeq
-\frac{M}{2}
\ ,
\label{calEn}
\ee
which corresponds to
\be
E_n^2
\ge
E_{N_M}^2
\simeq
0
\ .
\ee
Since any semiclassical state representing the collapsing ball must be described
in terms of superpositions of the $\Psi_n$'s with $n\ge N_M$, the above quantum bound~\eqref{eqX}
implies that any ball of dust must have compactness $\gn\,M/R\lesssim 1$.
It is important to stress once more that this result holds up to a factor of order one and better estimates
should be obtained only by considering more realistic and complete models.
However, it appears rather unnatural that such refinements can bring down the minimum size from a fraction
of $\Rh$ to $\lp$.
\begin{figure}[t]
\centering
\includegraphics[width=7cm]{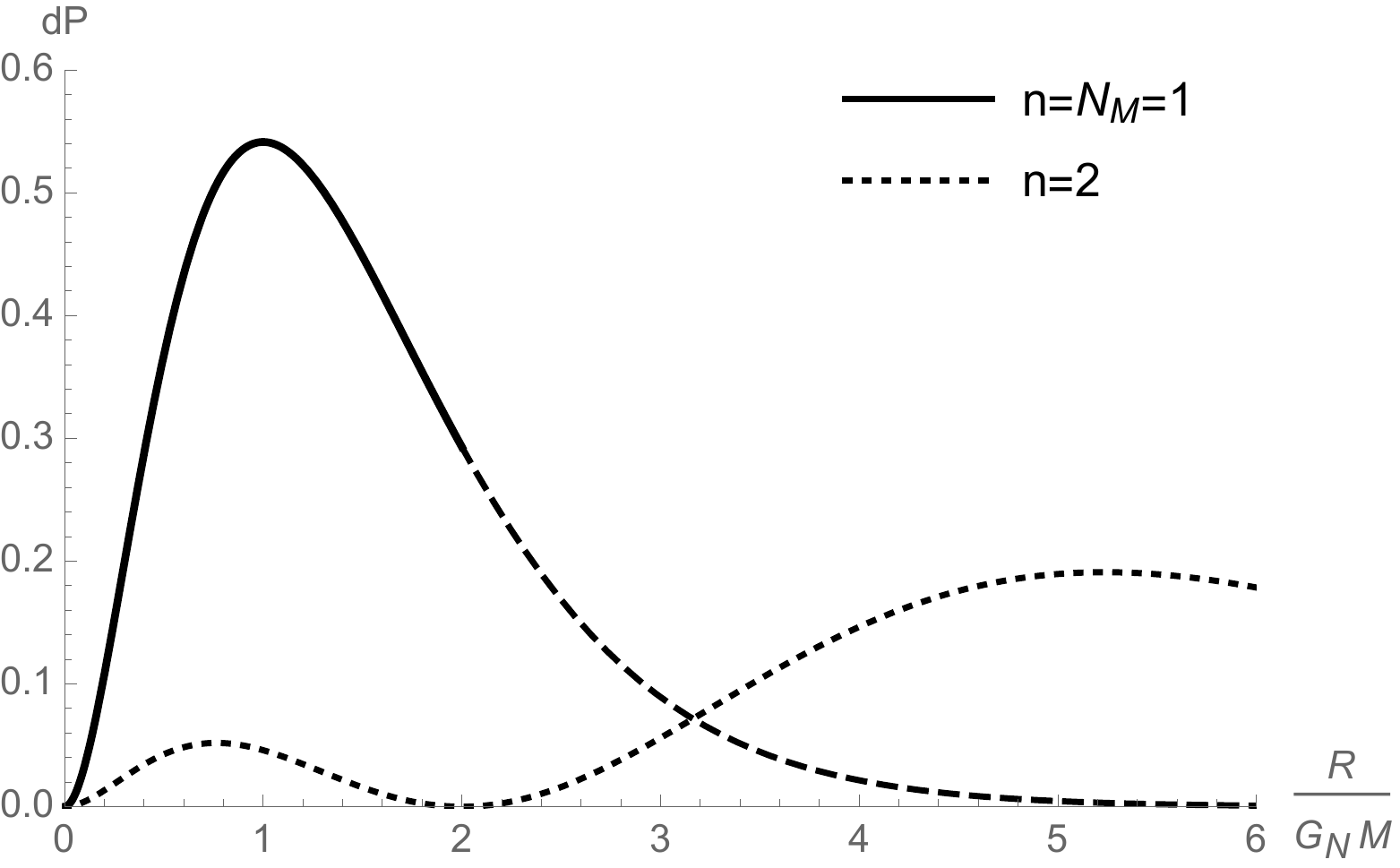}
$\qquad $
\includegraphics[width=7cm]{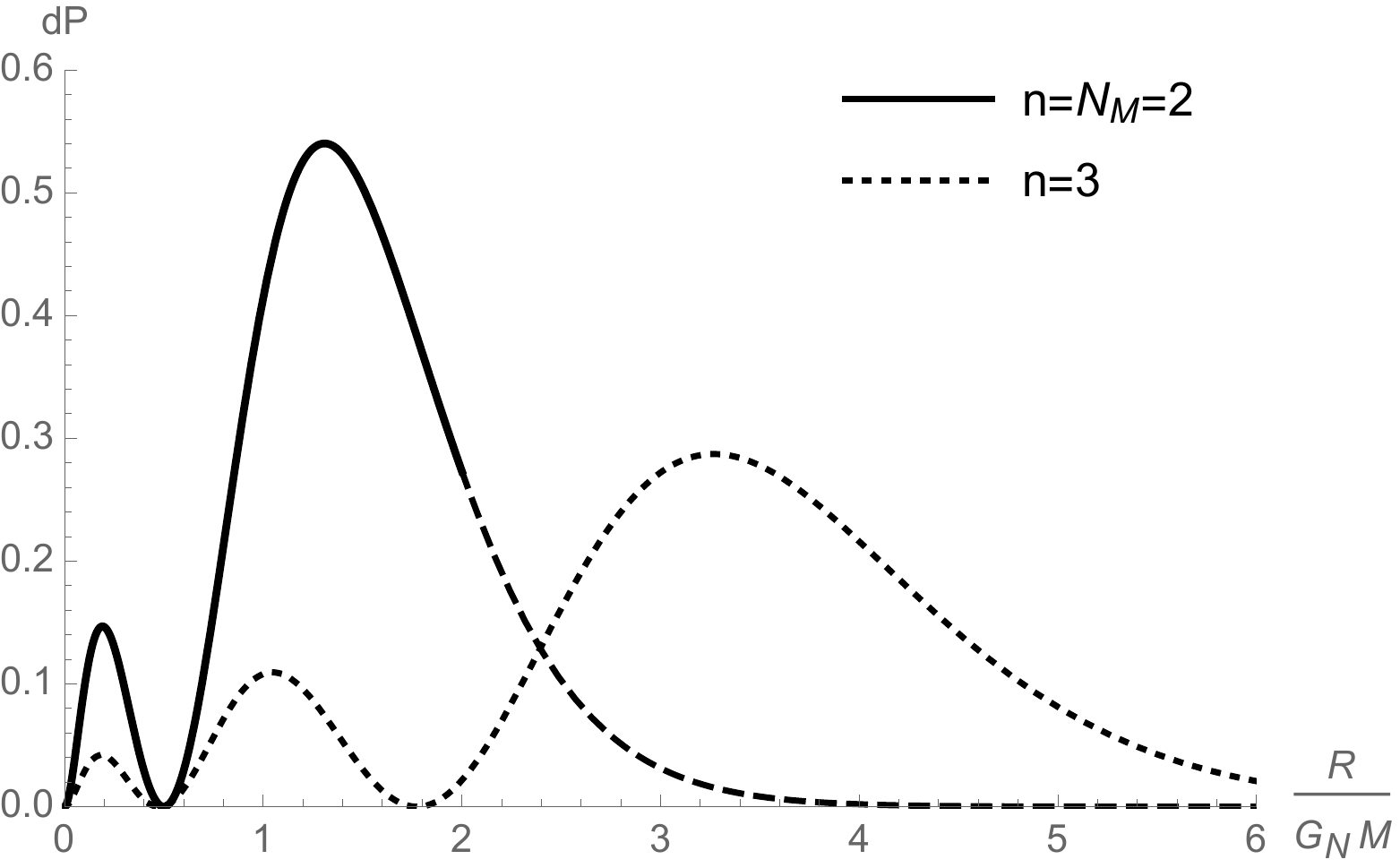}
\\
$\ $
\\
\includegraphics[width=7cm]{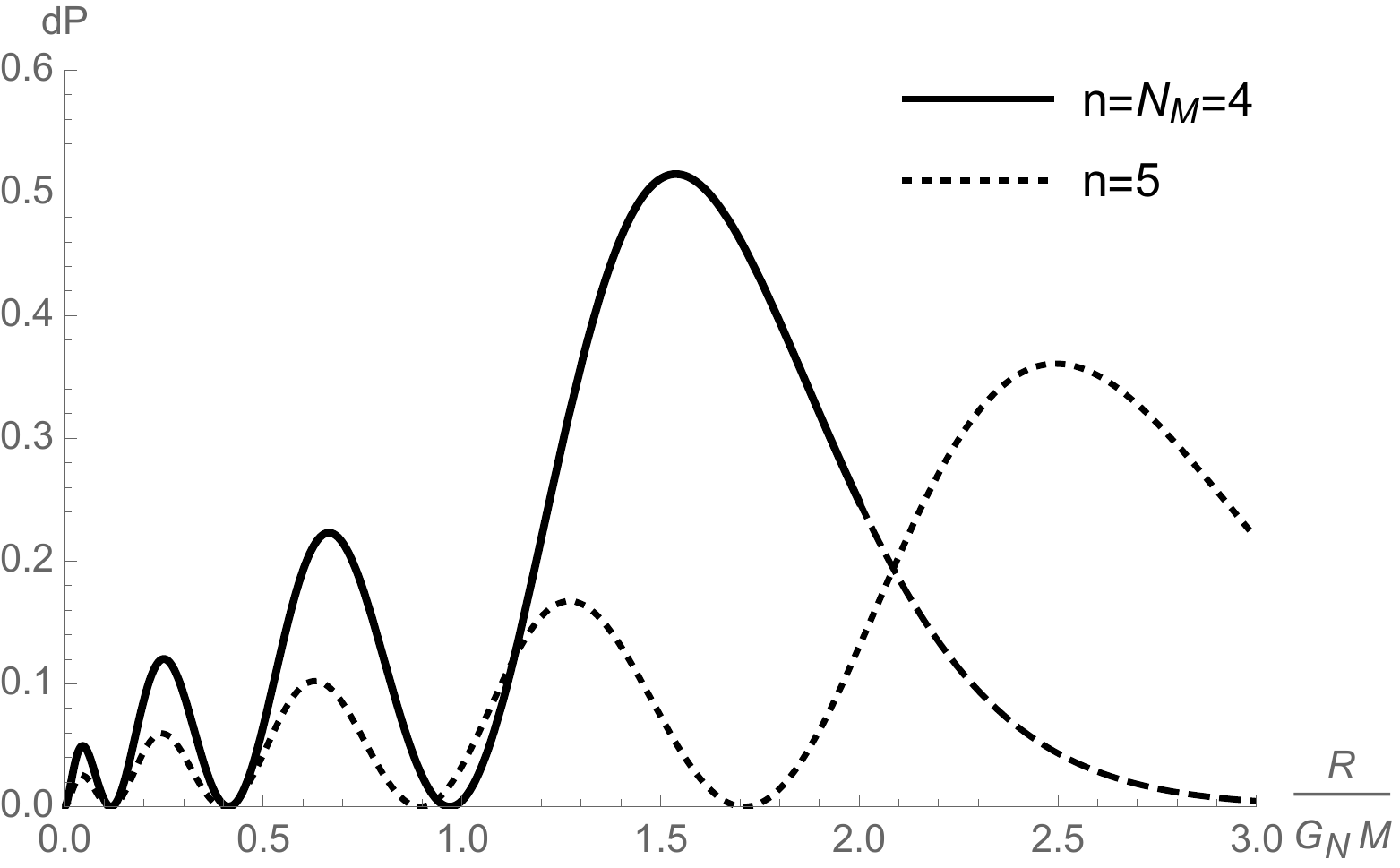}
$\qquad $
\includegraphics[width=7cm]{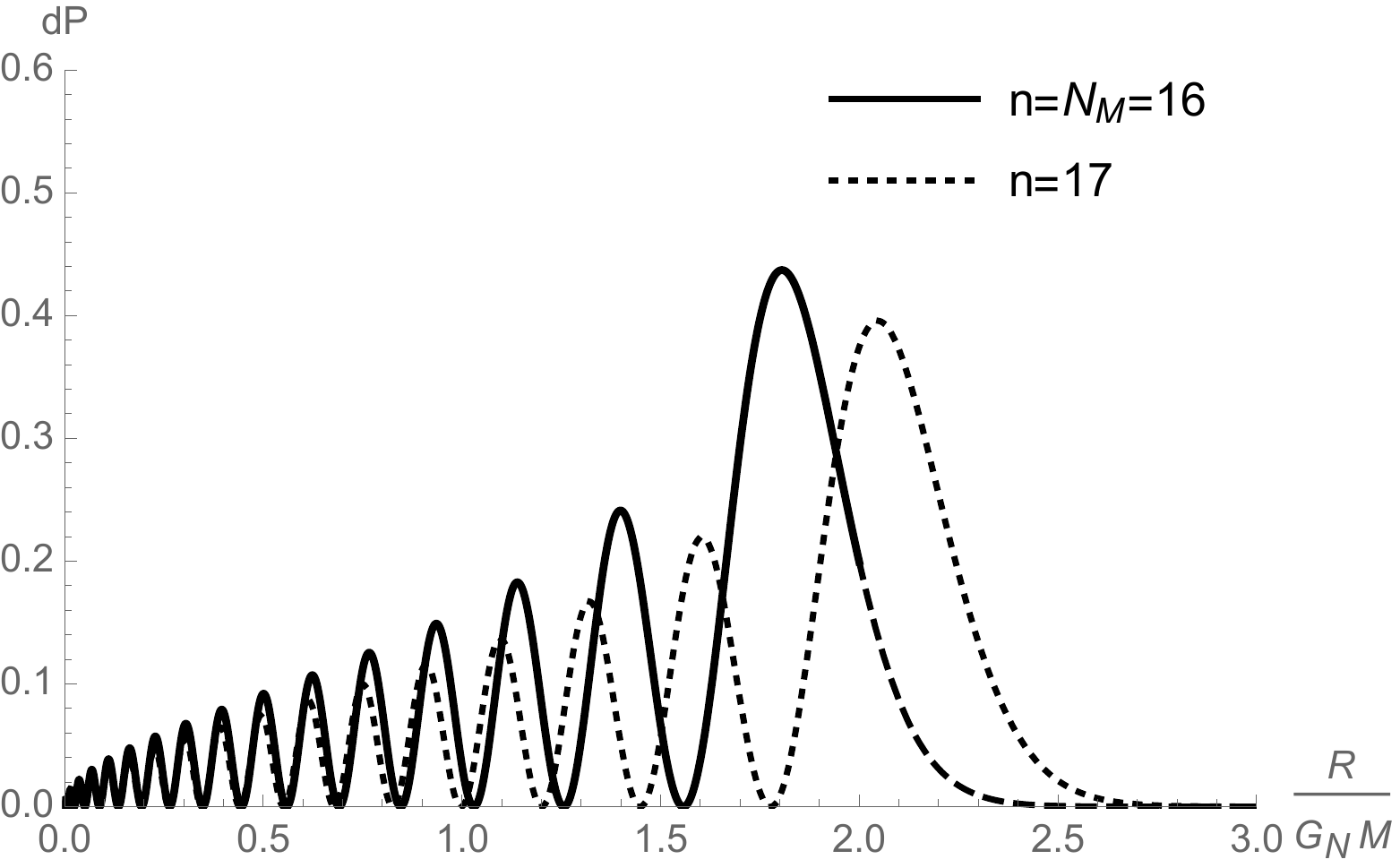}
\caption{Probability density $\mathcal P=\mathcal P(R)$ for: a) ground state with $n=N_M=M^2/\mpl^2$
[solid (dashed) line represents the region inside (outside) the gravitational radius $\Rh=2\,\gn\,M$];
b) first excited state with $n=N_M+1$ (dotted line).}
\label{dPNM}
\end{figure}
\par
The present minisuperspace description only contains the observable $R$.
From the wavefunction $\Psi=\Psi(R)$, we can therefore only determine such information as
the expectation value of $R$ and the probability that the ball be inside the gravitational radius,
\be
P(R\le \Rh)
\equiv
\int_0^{\Rh}
\mathcal P(R)\,\d R
=
4\,\pi
\int_0^{\Rh}
|\Psi(R)|^2\,
R^2\,\d R
\ ,
\ee
which can be viewed as the probability that the dust ball is a black hole (when the mass $M$ is
treated as a fixed parameter~\footnote{Alternative viewpoints are considered in Refs.~\cite{hqm},
where the mass $M$ is quantised, and in Ref.~\cite{mentrelli}, where $M$ is allowed 
a statistical uncertainty.
We will comment again about the former approach in Section~\ref{S:conc}.}).
For the ground state, whose wavefunction is given by 
\be
\Psi
=
\Psi_{N_M}
\simeq
\sqrt{\frac{\mpl}{\pi\,\lp^3\,M}}\,
e^{-\frac{M\,r}{\mpl\,\lp}}\,
L_{\frac{M^2}{\mpl^2}-1}^1
\left(\frac{2\,M\,r}{\mpl\,\lp}
\right)
\ ,
\label{PsiN}
\ee
the probability density $\mathcal P=\mathcal P_{N_M}(R)$ is plotted in Fig.~\ref{dPNM}
for a few cases with very small $M$.
For values of $M\gg \mpl$, the probability density narrowly peaks around a value of $R$
slightly below $\Rh$ and one thus finds 
\be
P_{N_M}(R\le \Rh)\simeq 1
\ .
\label{Pbh}
\ee
Moreover, the width of this highest peak 
\be
\Delta R_{N_M}
\sim
\frac{R_{N_M}}{N_M}
\sim
\lp\,\frac{\mpl}{M}
\ .
\label{dRbh}
\ee
As expected, $\Delta R_{N_M}\ll R_{N_M}$, hence the radius of a very massive matter source
should behave like a classical variable.
\par
Fig.~\ref{dPNM} also displays the first excited state, which asymptotically approaches the 
corresponding ground state for increasing $M$. 
This implies that $P_{n}(R\le \Rh)\simeq 1$ for a range of states $n\gtrsim N_M$,
which suggests that large astrophysical black holes of a given mass $M$ might
not necessarily be in their ground state with $n=N_M$. 
In fact, for $n\gtrsim N_M$, the quantum of the Hamiltonian $H$ in Eq.~\eqref{radial} is given by
\be
\delta H
\equiv
\left|\mathcal E_{n+1}
-
\mathcal E_{n}
\right|
\simeq
\mpl\,\frac{\mpl}{M}
\ ,
\label{dH}
\ee
so that $\delta H\sim \mpl\,(\Delta R_{N_M}/\lp)\ll \mpl$ for a macroscopic object of mass $M\gg \mpl$.
Furthermore, since
\be
\delta E
\equiv
\left|E_{n+1}
-
E_{n}\right|
\simeq
\mpl
\ ,
\ee
it appears that the proper source ``energy'' is naturally quantised in units of the fundamental
Planck mass $\mpl$, but this quantum is redshifted down to the much smaller $\delta H$ 
measured by outer observers.
\section{Concluding remarks and outlook} 
\label{S:conc}
\setcounter{equation}{0}
We have shown how the General Relativistic description of gravity leads to
an upper bound for the compactness of a ball of dust in the quantum theory.
It is interesting to notice that the lower bounds on the eigenvalue $\mathcal E_n\ge \mathcal E_{N_M}$
and radius $R_n\gtrsim \gn\,M$ could not be obtained in the Newtonian theory, since $\mathcal E=E$
in that approximation, and the entire spectrum $\Psi_n$ with $n\ge 1$ would be physically acceptable therein.
The bound~\eqref{eqX} on the compactness therefore follows from the nonlinearity
of General Relativity and agrees with previous results~\cite{Casadio:2020ueb} obtained by adding
a gravitational self-interaction term to the Newtonian theory~\cite{BootN}.
It also agrees with those results following from the quantum description of the gravitational radius and
black hole horizon~\cite{hqm}.
In particular, the latter approach leads to very similar conclusions to the ones shown in the previous 
Section, like the black hole probability~\eqref{Pbh} and radius uncertainty~\eqref{dRbh}, when the
self-gravitating object is described by an extended many-body system with a very large occupation
number of order $N_M\sim M^2/\mpl^2$ in its ground state~\cite{Casadio:2015bna}.
The first excited modes could then be populated thermally~\cite{Casadio:2015bna} and reproduce
the Hawking radiation~\cite{hawking}.
\par
It is indeed very intriguing that Eq.~\eqref{eqn0} and the bound~\eqref{eqX} 
allow for recovering the fundamental scaling relations employed in the corpuscular description
of black holes~\cite{DvaliGomez}.
In fact, the allowed state of minimum energy~\eqref{PsiN} has principal quantum number 
\be
N_M
\simeq
N_{\rm G}
\ ,
\ee
where $N_{\rm G}$ is the number of soft gravitons in the coherent state representing
the gravitational potential generated by a source of mass $M$~\cite{coherent,Casadio:2021eio}.
We recall that the corpuscular picture naturally reproduces Bekenstein's area
spectrum $\gn^2\,M^2\sim N_{\rm G}\,\lp^2$~\cite{bekenstein}, so that the black hole ADM mass
is quantised in units of
\be
\delta M
\simeq
\delta H
\ ,
\label{dM}
\ee
where $\delta H$ is precisely the quantum of our Hamiltonian given in Eq.~\eqref{dH}.
This is also the typical energy of particles emitted by the Hawking evaporation process,
which the corpuscular picture describes as the depletion of the quantum state of gravity~\cite{DvaliGomez}.
\par
In this perspective, the wavefunction $\Psi_{N_M}=\Psi_{N_{\rm G}}$ appears as the
``non-perturbative ground state'' for self-gravitating macroscopic objects of mass $M$
and should thus be the closest possible to a classical black hole (as we have already commented
at the end of the previous Section).
This result is consistent with the fact that a static gravitational field must be fully determined by the
state of the source, as it does not contain independent degrees of freedom in a static
system~\cite{hqm}.
It can further be interpreted as the fact that the quantum state of a macroscopic self-gravitating
system of mass $M$ is very far from the vacuum $\Psi_0$ of quantum gravity, for which 
$N_{\rm G}=0$.
The number $N_M\sim N_{\rm G}\sim M^2$ hence provides a quantitative measure
for this ``distance'' from the vacuum in the Hilbert space of quantum gravity states.
The appearance of the ground state $\Psi_{N_M}=\Psi_{N_{\rm G}}$ when the ADM energy is
given by Eq.~\eqref{eqn0}, in turn, can be interpreted as a form of {\em classicalization\/}
ensuring the ultraviolet self-completeness of gravity~\cite{classicalization}, because states with spatial
momenta much larger than $\hbar/R_{N_M}\sim \hbar/\Rh$ have $n<N_M$ and cannot be populated
at energy scales of the order of the mass $M$.
\par
We note that, although quantum states corresponding to the classical singularity appear to be removed
from the spectrum, it would be phenomenologically very important to determine a more precise
maximum value of the compactness in order to assess the size of quantum deviations from the exterior 
Schwarzschild geometry.
Moreover, in order to make contact with actual observations, one needs an explicit description of
the exterior spacetime and of the interaction among the collapsing object and signals 
that can reach out to our detectors.
For example, an effectively finite size of the self-gravitating system is expected
to induce the kind of quantum deviations obtained in Ref.~\cite{Casadio:2021eio}
from a ultraviolet cut-off for the momenta generating the outer mean-field geometry. 
A unified description of the quantum states of the collapsing ball presented
here and the quantum gravitational potential of Ref.~\cite{Casadio:2021eio} 
is left for future investigations.
\par
We conclude by remarking that spinning objects and electrically charged sources would also be
very interesting to consider.
In particular, charged sources have already been analysed by means of the horizon quantum
mechanics~\cite{hqm}, for which it was then shown that the inner Cauchy horizon of the classical
Reissner-Nordstr\"om black hole has vanishing probability to occur~\cite{Q1} and that quantum fluctuations
do not allow for the existence of an event horizon if the source is (significantly) overcharged~\cite{Q2}.
The case of rotating black holes was likewise analysed in Ref.~\cite{Casadio:2017nfg}, and the existence
of the inner Cauchy horizon of the Kerr geometry was again found to be highly disfavoured,
although the angular momentum could be included only as an asymptotic quantity.
We expect the present approach will lead to similar conclusions, although more preliminary work
is necessary in order to describe the collapse of rotating or electrically charged objects by means
of a manageable minisuperspace.
\section*{Acknowledgments}
R.C.~is partially supported by the INFN grant FLAG and his work has also been carried out in
the framework of activities of the National Group of Mathematical Physics (GNFM, INdAM)
and COST action {\it Cantata\/}. 

\begin{thebibliography}{99}
%
%
\bibitem{HE}
S.~W.~Hawking and G.~F.~R.~Ellis,
``The Large Scale Structure of Space-Time''
(Cambridge University Press, Cambridge, 1973)
%
\bibitem{OS}
J.~R.~Oppenheimer and H.~Snyder,
Phys. Rev. \textbf{56} (1939) 455.
%
\bibitem{ADM}
R.L.~Arnowitt, S.~Deser and C.W.~Misner,
Phys.\ Rev.\  {\bf 116} (1959) 1322.
%
\bibitem{bounce}
V.~P.~Frolov and G.~A.~Vilkovisky,
Phys. Lett. B \textbf{106} (1981) 307;
D.~Malafarina,
Universe \textbf{3} (2017) 48
[arXiv:1703.04138 [gr-qc]];
R.~Casadio,
Int. J. Mod. Phys. D \textbf{9} (2000) 511
[arXiv:gr-qc/9810073 [gr-qc]];
H.~M.~Haggard and C.~Rovelli,
Phys. Rev. D \textbf{92} (2015) 104020
[arXiv:1407.0989 [gr-qc]];
W.~Piechocki and T.~Schmitz,
Phys. Rev. D \textbf{102} (2020) 046004
[arXiv:2004.02939 [gr-qc]];
T.~Schmitz,
Phys. Rev. D \textbf{103} (2021) 064074
[arXiv:2012.04383 [gr-qc]].
%
\bibitem{stephani}
H.~Stephani,
``Relativity''
(Cambridge University Press, Cambridge, 2004)
%
\bibitem{qcollapse}
P.~Hajicek, B.~S.~Kay and K.~V.~Kuchar,
Phys. Rev. D \textbf{46} (1992) 5439;
P.~Hajicek and C.~Kiefer,
Int. J. Mod. Phys. D \textbf{10} (2001) 775
[arXiv:gr-qc/0107102 [gr-qc]];
R.~Casadio and G.~Venturi,
Class. Quant. Grav. \textbf{13} (1996) 2715
[arXiv:gr-qc/9512032 [gr-qc]].
%
\bibitem{hqm}
R.~Casadio,
``Localised particles and fuzzy horizons: A tool for probing Quantum Black Holes,''
[arXiv:1305.3195 [gr-qc]];
R.~Casadio and F.~Scardigli,
Eur. Phys. J. C \textbf{74} (2014) 2685
[arXiv:1306.5298 [gr-qc]];
R.~Casadio, A.~Giugno, O.~Micu and A.~Orlandi,
Phys. Rev. D \textbf{90} (2014) 084040
[arXiv:1405.4192 [hep-th]];
R.~Casadio, A.~Giugno and A.~Giusti,
Gen. Rel. Grav. \textbf{49} (2017) 32
[arXiv:1605.06617 [gr-qc]].
%
\bibitem{mentrelli}
R.~Casadio, A.~Giusti and A.~Mentrelli,
Phys. Rev. D \textbf{100} (2019) 024036
[arXiv:1901.06206 [gr-qc]].
%
\bibitem{Casadio:2020ueb}
R.~Casadio, M.~Lenzi and A.~Ciarfella,
Phys. Rev. D \textbf{101} (2020)124032
[arXiv:2002.00221 [gr-qc]].
%
\bibitem{BootN}
R.~Casadio, M.~Lenzi and O.~Micu,
  Phys.\ Rev.\ D {\bf 98} (2018) 104016
  [arXiv:1806.07639 [gr-qc]];
%
   Eur.\ Phys.\ J.\ C {\bf 79} (2019) 894
  [arXiv:1904.06752 [gr-qc]];
R.~Casadio and I.~Kuntz,
Eur. Phys. J. C \textbf{80} (2020) 581
[arXiv:2003.03579 [gr-qc]];
R.~Casadio, A.~Giusti, I.~Kuntz and G.~Neri,
Phys. Rev. D \textbf{103} (2021) 064001
[arXiv:2101.12471 [gr-qc]].
%
\bibitem{Casadio:2015bna}
R.~Casadio, A.~Giugno and A.~Orlandi,
Phys. Rev. D \textbf{91} (2015) 124069
[arXiv:1504.05356 [gr-qc]];
%
\bibitem{hawking}
S.~W.~Hawking,
Commun. Math. Phys. \textbf{43} (1975) 199
[erratum: Commun. Math. Phys. \textbf{46} (1976) 206].
%
\bibitem{DvaliGomez} 
G.~Dvali and C.~Gomez,
Fortsch.\ Phys.\  {\bf 61} (2013) 742
[arXiv:1112.3359 [hep-th]];
G.~Dvali, C.~Gomez and S.~Mukhanov,
``Black Hole Masses are Quantized,''
arXiv:1106.5894 [hep-ph].
G.~Dvali and C.~Gomez,
Phys.\ Lett.\ B {\bf 719} (2013) 419
[arXiv:1203.6575 [hep-th]];
Phys.\ Lett.\ B {\bf 716} (2012) 240
[arXiv:1203.3372 [hep-th]];
Eur.\ Phys.\ J.\ C {\bf 74} (2014) 2752
[arXiv:1207.4059 [hep-th]];
S.~Hofmann and T.~Rug,
Nucl. Phys. B \textbf{902} (2016) 302
[arXiv:1403.3224 [hep-th]];
A.~Giusti,
Int.\ J.\ Geom.\ Meth.\ Mod.\ Phys.\  {\bf 16} (2019) 1930001.
%
\bibitem{coherent}
R.~Casadio, A.~Giugno and A.~Giusti,
Phys.\ Lett.\ B {\bf 763} (2016) 337
[arXiv:1606.04744 [gr-qc]];
%
W.~M\"uck,
Can. J. Phys. \textbf{92} (2014) 973
[arXiv:1306.6245 [hep-th]];
%
R.~Casadio, A.~Giugno, A.~Giusti and M.~Lenzi,
Phys.\ Rev.\ D {\bf 96} 044010 (2017)
[arXiv:1702.05918 [gr-qc]].
%
\bibitem{Casadio:2021eio}
R.~Casadio,
``Quantum black holes and resolution of the singularity,''
[arXiv:2103.00183 [gr-qc]].
%
\bibitem{bekenstein}
J.~D.~Bekenstein,
Phys. Rev. D \textbf{7} (1973) 2333.
%
\bibitem{classicalization}
G.~Dvali, G.~F.~Giudice, C.~Gomez and A.~Kehagias,
JHEP {\bf 1108} (2011) 108
[arXiv:1010.1415 [hep-ph]];
G.~Dvali and D.~Pirtskhalava,
  Phys.\ Lett.\ B {\bf 699} (2011) 78
  [arXiv:1011.0114 [hep-ph]];
  G.~Dvali, C.~Gomez and A.~Kehagias,
  JHEP {\bf 1111} (2011) 070
  [arXiv:1103.5963 [hep-th]];
R.~Percacci and L.~Rachwal,
  Phys.\ Lett.\ B {\bf 711} (2012) 184
  [arXiv:1202.1101 [hep-th]].
%
\bibitem{Q1}
R.~Casadio, O.~Micu and D.~Stojkovic,
JHEP \textbf{05} (2015) 096
[arXiv:1503.01888 [gr-qc]].
%
\bibitem{Q2}
R.~Casadio, O.~Micu and D.~Stojkovic,
Phys. Lett. B \textbf{747} (2015) 68
[arXiv:1503.02858 [gr-qc]].
%
\bibitem{Casadio:2017nfg}
R.~Casadio, A.~Giugno, A.~Giusti and O.~Micu,
Eur. Phys. J. C \textbf{77} (2017) 322
[arXiv:1701.05778 [gr-qc]].
%
\end{thebibliography}
\end{document}